\documentclass{elsart}

\usepackage{natbib}

\usepackage{epsfig}

\usepackage{amssymb}

\begin{document}

\begin{frontmatter}



\title{Modeling of waiting times and price changes in currency exchange data.}
\author{Przemys\mbox{\l}aw Repetowicz }
\ead{repetowp@tcd.ie}
\author{Peter Richmond}
\ead{richmond@tcd.ie}
\address{Department of Physics, Trinity College Dublin 2, Ireland}

\begin{abstract}
A theory which describes the share price evolution 
at financial markets as a continuous-time random walk 
\citet{ScalGor,MainRaber,SabatKeat,RaberScal} has been generalized 
in order to take into account the dependence of waiting times $t$ on price returns $x$.
A joint probability density function (pdf) $\phi_{X,T}(x,t)$ which uses
the concept of a L\'{e}vy stable distribution is worked out.
The theory is fitted to high-frequency 
US \$/Japanese Yen  exchange rate and low-frequency 19th century Irish stock data. 
The theory has been fitted both to price return and to waiting time data
and the adherence to data, in terms of the $\chi^2$ test statistic,
has been improved when compared to the old theory 
\citet{SabatKeat}.
\end{abstract}


\end{frontmatter}

\section{Introduction}
\label{sec:intro}
The idea to describe the stock market share prices as a random walk
dates back to \citet{Bach}. 
He assumed that price returns $x$ and waiting times $t$
are independent, identically distributed (i.i.d.) random variables
which in the long time limit, according to the Central Limit Theorem (CLT),
conform to the normal distribution.

The properties of random walks are well understood [see for example \citet{Montr}]
and since low-frequency market data conforms fairly well to a normal distribution
the concept of a Continuous Time Random Walk (CTRW) 
has been incorporated into 
standard economical textbooks. This approach enabled \citet{Black}
to formulate 
a theory of pricing financial 
derivatives for which they received the Nobel price in economics.

However there are serious problems with CTRWs. 
The observed probability density function (pdf) of price returns $\phi(x)$ 
shows considerable 
discrepancies from a Gaussian [\citet{Lux,GopiPle,PleGopi}]
In particular it exhibits power law tails for large values of $x$. 
\citet{Mandel} and \citet{Fama} managed to generalize 
Bachelier's approach by introducing  L\'{e}vy distributions $L_{\mu,\beta}(x)$
with the normal distribution being only 
a particular representative of these.

L\'{e}vy distributions, functions labeled by two real parameters $\alpha$ and $\beta$,
are limit distributions for sums of independent random variables 
which emerge in the generalized CLT \citet{Bouch}.
Taking $\beta = 0$ and $0 < \alpha < 2$ ensures that the functions are both positive, even 
$\left( L_{\mu,\beta}(-x) = L_{\mu,\beta}(x) \right)$, 
behave like a steep Gaussian for small values of $x$ and decay 
according to a Pareto's power law much slower than a Gaussian for large $x$'s.
Despite technical problems, the second moment being infinite for $\alpha < 2$,
these functions have been successfully used to model relaxation processes
(die-electrical, mechanical and NMR)
\citet{MontrBendl} and in the theory of probability \citet{Levy,Gnedenko}.

Recently, an
investigation of about 40 million price quotes at the New York Stock Exchange (NYSE)
\citet{GopiMeyer} showed that the cumulative probability density of price returns $S(x) = P(X>x)$ behaves like
$S(x) \sim x^{-\mu}$ with $\mu = 3.1 \pm 0.003$ which yields a value of $\mu$ beyond
the valid range $0 < \mu \le 2$ if the pdf were to be modeled by a L\'{e}vy distribution.
L\'{e}vy distributions are well defined for $\mu > 2$ but can take negative values
in that regime hence are not good candidates for pdfs.

Furthermore in most treatments so far it is 
assumed that price returns $x$
are independent of waiting times $t$. 
This does not seem to be the case \citet{RaberScal} and it also seems counterintuitive.
One would expect larger price variations to correspond to larger waiting times
since one would expect it to be more difficult, take more time, for a broker to 
find buyers and sellers that balance out supply and demand
if the price variation is larger than if it was smaller.
Such simplified reasoning implies that there should be a positive correlation
between $x$'s and $t$'s which has not been taken into account in many works on this subject.

In this paper we will generalize the theory to take account of these two difficulties.
We will drop the assumption that $x$'s and $t$'s are independent random variables but still retain
the property of independence of variables corresponding to different times.
By assuming that it is a certain function of $x$ and $t$ that conforms asymptotically 
to the L\'{e}vy distribution
rather than either price returns or waiting times themselves we will work out the joint pdf $\phi_{X,T}(x,t)$.
Theoretical results will be compared to some high-frequency market data.

The paper is organized as follows.
In section \ref{sec:theory} we formulate our theory and work out analytical formulas 
for pdfs and cumulative pdfs under various assumptions about how price returns
and waiting times are correlated. In section \ref{sec:masterEq} we set up an equation 
for the probability $p(x, t)$ of having price $x$ at time $t$ and show that 
price evolution does not satisfy the Markovian property. We compare our
evolution equation to that worked out under different assumptions \cite{SabatKeat}.
Finally in sections \ref{sec:CumulDistr}, \ref{sec:GoodFit} and \ref{sec:Correl} we fit
the cumulative pdfs of
high-frequency exchange rate and low-frequency $19$th century stock data to our theory,
discuss the goodness of fit and assess whether the correlations between $x$'s and $t$'s
which follow from our theory conform to market data.

\section{A random walk with non-independent price returns and waiting times.}
\subsection{Pdfs and cumulative pdfs. \label{sec:theory}}
In the following we will adhere to the convention that random variables are 
denoted by capital letters and values of random variables by lower-case letters.
The objective is to work out the joint pdf of price returns and waiting times 
$\phi_{X,T}(x,t)$ under the assumption that the random variables $X$ and $T$ 
are not independent.
The first idea which comes into a mind of a physicist is to assume a weak dependence and to develop 
a perturbative approach.

One could for example consider the correlation coefficient between $X$ and $T$:
\begin{equation}
\mbox{corr($X,T$)} = \frac{E[XT] - E[X]E[T]}{\sqrt{E[X^2] - E[X]^2} \sqrt{E[T^2] - E[T]^2}}
\end{equation}
with $E[f]$ being the mean of $f$ 
and minimize it as a functional of the joint pdf $\phi_{X,T}(x,t)$ subject to an assumption that
the pdf belongs to a certain functional space, ie is a linear combination of certain basis functions.
However, since it is rather unclear how to choose these functions we will develop a different approach.

We assume that it is not the price return $x$ or the waiting time $t$ that conforms to a L\'{e}vy distribution
but that a certain function $f_t = f(x(t),t)$ has that property. 
This means that there exists a random variable $F_T$, which depends on the random price returns $X$ and waiting times $T$, 
the change of which in the time period $[T_1, T_N]$
takes, for large $N$, the following form:
\begin{equation}
\frac{F_{T_N} - F_{T_1}}{N} = \frac{\sum\limits_{i=1}^N F_{T_{i+1}} - F_{T_i}}{N}
                                     = \left< F \right> + \frac{1}{N^{1 - 1/\alpha}} \;\cdot\; \mbox{L\'{e}vy($\alpha,\beta$)}
\label{eq:GenCLT}
\end{equation}
where $\mbox{L\'{e}vy($\alpha,\beta$)}$ is a random variable conforming to a L\'{e}vy distribution with parameters $\alpha$ and $\beta$.
Equation (\ref{eq:GenCLT}) is a statement of the Generalized CLT under the assumption that the increments
$F_{T_{i+1}} - F_{T_i}$ are i.i.d random variables.
We also assume
that instead of one stock there can be a set of stocks $\vec{X} = \{X_1,\ldots,X_p\}$ the prices of which are 
correlated with each other and so $F_T$ depends on all stock price returns $\vec{X}$, ie $F_T = F(\vec{X},T)$,
and $F_T$ conforms to the L\'{e}vy distribution (\ref{eq:GenCLT}).

In order to work out the distribution $\phi_{X,T}(x,t)$ of the returns and waiting times 
we consider a transformation of variables in the $p+1$-dimensional space with Cartesian 
coordinates $x_1,\ldots,x_p,t$:
\begin{equation}
 \left( x_1,x_2,\ldots,x_p,t \right) \longrightarrow \left( f, \theta_1,\ldots,\theta_p \right) 
\label{eq:mapping}
\end{equation}
We require the elementary probabilities $P\left( \right)$ to be conserved:
\begin{equation}
	 L_{\mu,0}(f) df \Theta_1(\theta_1)d\theta_1\ldots \Theta_p(\theta_p)d\theta_p = 
        	   P\left(\; \begin{array}{c} 
			      f \le F \le f+df\\
			      \theta_i \le \Theta_i \le \theta_i+d\theta_i 
			     \end{array}        \right) \;:=\; \phi(\vec{x},t)d\vec{x} dt
\label{eq:trafo}
 \end{equation}
where $\Theta_i(\theta)$ for $i=1,\ldots,p$ are certain functions of the variables $\theta_i$
which we set equal to one $\Theta_i = 1$.

\begin{figure}[!h]
\centerline{\psfig{figure=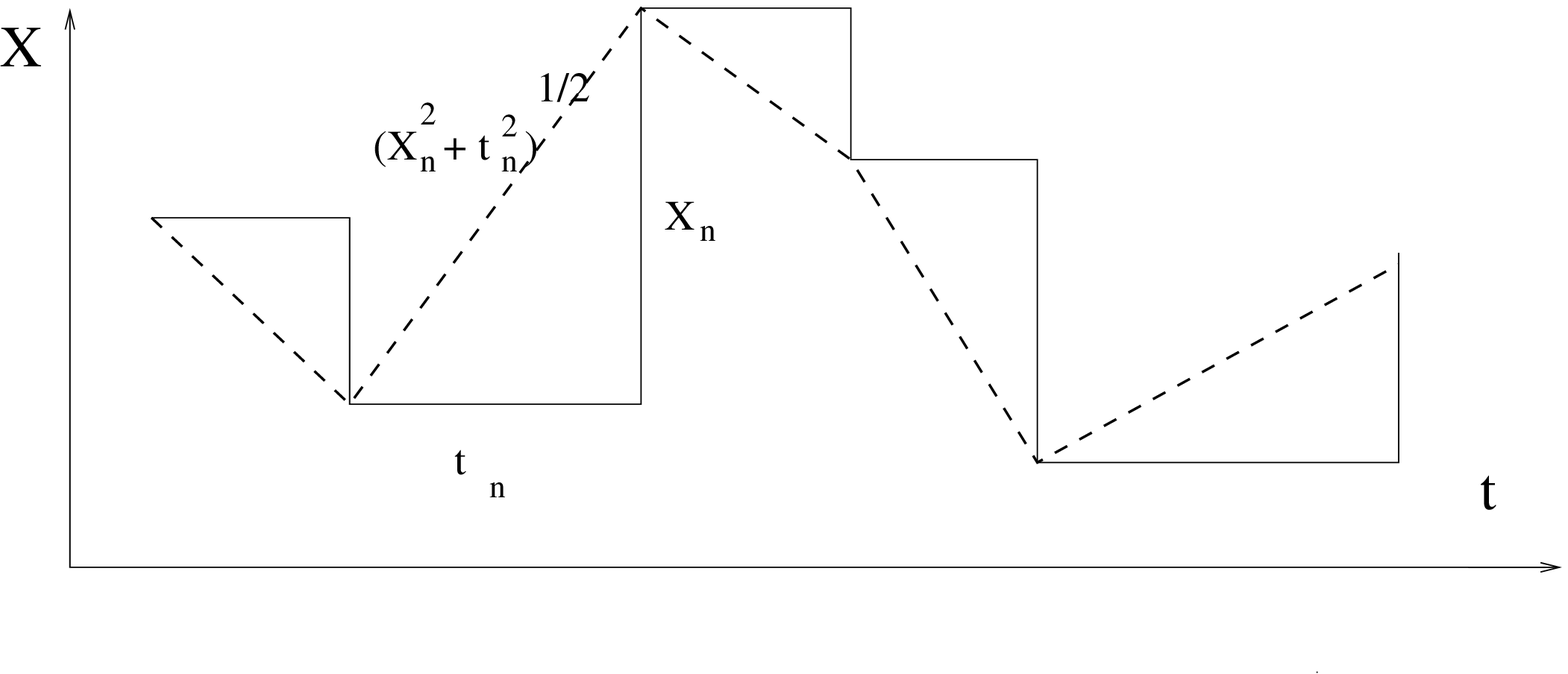,width=0.99\textwidth,angle=0}}
\caption{Illustration of our assumption concerning the random walk.
         Here the changes in lengths of consecutive hypotenuses(marked with dashed lines)
         are i.i.d. random variables. Therefore, on the grounds of the generalized CLT,
         the length of the hypotenuse conforms, in large time limit, to a L\'{e}vy distribution
         (compare (\ref{eq:GenCLT})). 
         \label{fig:RandWalk}}
\end{figure}

Now the problem is to make a proper choice of the function $f = f(\vec{x},t)$.
We will assume that $f$ is a hypotenuse in a right triangle with edges 
of lengths $|\vec{x}|$ and $t$ (see Fig.\ref{fig:RandWalk}). 
Therefore $f = \sqrt{ x_1^2 + \ldots + x_p^2 + t^2 }$ and the most natural 
transformation (\ref{eq:mapping}) that comes into mind here is a mapping into
spherical polar coordinates \citet{Arfken}: 
\begin{equation}
  \begin{array}{lll}
          x_1 & = & f \sin \theta_p \sin \theta_{p-1} \ldots \sin\theta_2 \sin\theta_1 \\
          x_2 & = & f \sin \theta_p \sin \theta_{p-1} \ldots \sin\theta_2 \cos\theta_1 \\
          x_3 & = & f \sin \theta_p \sin \theta_{p-1} \ldots \cos\theta_2  \\
          \vdots & = & \vdots \\
          x_p & = & f \sin\theta_p \cos\theta_{p-1} \\
          t   & = & f \cos\theta_p \\
  \end{array}
\label{eq:trafoI}
\end{equation}
where $0 \le \theta_p \le \pi$ and $0 \le \theta_j \le 2\pi$ for $j=1,\ldots,p-1$.
The Jacobian of the transformation (\ref{eq:trafoI}) reads:
\begin{equation}
         J = \left| \frac{\partial(x_1,\ldots,x_p,t)}{\partial(f,\theta_1,\ldots,\theta_p)} \right|  = 
          f^{p} \sin\theta_2 \sin^2\theta_3 \ldots \sin^{p-1}\theta_p =
          f \cdot {\mathcal M}(\vec{x}) 
\label{eq:Jacobian}
\end{equation}
where 
\begin{equation}
{\mathcal M}(\vec{x}) = \sqrt{\left( \prod_{q=2}^p (\sum_{s=1}^q x_s^2) \right)}
\label{eq:JacobianI}
\end{equation}
and the joint pdf of price returns and waiting times reads:
\begin{eqnarray}
\phi_{\vec{X},T}(\vec{x},t) &=& L_{\mu,0}(f) \cdot \left| 
\frac{\partial(f,\theta_1,\ldots,\theta_p)}{\partial(x_1,\ldots,x_p,t)} \right| \nonumber \\
&=& \frac{L_{\mu,0}(f)}{f} \frac{1}{ {\mathcal M}(\vec{x}) } 
 =  \frac{L_{\mu,0}(\sqrt{|\vec{x}|^2 + t^2})}{\sqrt{|\vec{x}|^2 + t^2}}
    \frac{1}{ {\mathcal M}(\vec{x}) } 
\end{eqnarray}

Integrating over all $\vec{x}$ values with a fixed modulus $x$ we obtain
the pdf $\phi_{X,T}(x,t)$ which corresponds to the length 
of the vector of price returns $x = |\vec{x}|$ and to the waiting time $t$ :
\begin{equation}
\phi_{X,T}(x,t) = \int \phi_{\vec{X},T}(\vec{x},t) \cdot 
\left| \frac{\partial(x_1,\ldots,x_p)}{\partial(x,\xi_1,\ldots,\xi_{p-1})} \right| d\xi_1 \ldots d\xi_{p-1} = \frac{2}{\pi} \frac{L_{\mu,0}(\sqrt{x^2 + t^2})}{\sqrt{x^2 + t^2}} 
\label{eq:JointPDF}
\end{equation}
In equation (\ref{eq:JointPDF}) we applied a mapping into spherical polar coordinates 
in $p$-dimensional space 
$\left( x_1, \ldots, x_p \right) \longrightarrow \left( x, \xi_1, \ldots, \xi_{p-1} \right)$
the Jacobian of which, according to equation (\ref{eq:Jacobian}), reads:
\begin{eqnarray}
\lefteqn{
\left| \frac{\partial(x_1,\ldots,x_p)}{\partial(x,\xi_1,\ldots,\xi_{p-1})} \right| 
= x \cdot \sqrt{\left( \prod_{q=2}^{p-1} (\sum_{s=1}^q x_s^2) \right)} =
} \nonumber \\
&& x \cdot \sqrt{ x^2 - x_p^2 } \cdot \sqrt{x^2 - x_p^2 - x_{p-1}^2} \cdot \ldots \cdot \sqrt{ x_1^2 + x_2^2}
= {\mathcal M}(\vec{x})
\end{eqnarray}

Finally we work out the cumulative density of waiting times \\
$S_T(t) = P( T > t )$
and that of price returns $S_X(x) = P( X > x )$:
\begin{eqnarray}
 S_T(t)  &=& \int\limits_t^\infty \phi_T(\xi) d\xi  
          = \frac{2}{\pi} \int\limits_t^\infty \int\limits_0^\infty \frac{L_{\mu,0}(\sqrt{\eta^2 + \xi^2})}{\sqrt{\eta^2 + \xi^2}} d\eta d\xi 
\label{eq:CumulDistr1}
\end{eqnarray}
Now, the double integral on the right-hand-side in (\ref{eq:CumulDistr1}) is computed by
going to polar coordinates $\eta = z\cos{\phi}$ and $\xi = z\sin{\phi}$.
The cumulative density $S_T(t)$ reads:
\begin{eqnarray}
S_T(t)   &=& \frac{2}{\pi} \int\limits_t^\infty dz z \frac{L_{\mu,0}(z)}{z} \int_{\phi_0}^{\pi/2} d\phi
         = \frac{2}{\pi} \int\limits_t^\infty dz L_{\mu,0}(z) \cdot 
           \left( \pi/2 - \phi_0 \right) \nonumber \\
         &=& \frac{2}{\pi} \int\limits_t^\infty 
                 \arccos{\left(\frac{t}{z}\right)} L_{\mu,0}(z) dz 
\label{eq:waitingTimedistrfct}
\end{eqnarray}
where $\phi_0 = \arcsin{(t/z)}$.
Since the marginal pdfs $\phi_X(x)$ and $\phi_T(t)$ take the same form,
it is readily seen that the joint pdf in (\ref{eq:JointPDF}) does not 
change when $x$ and $t$ are mutually exchanged,
the cumulative density of returns $S_X(x)$ is given by the same formula
as (\ref{eq:waitingTimedistrfct}) except that $t$ is replaced by $x$.
Let us also write down the large $t$ expansion of the cumulative density.
\begin{equation}
 S_T(t) = \frac{1}{2\pi\mu} \sum_{k=1}^\infty \frac{(-1)^{k+1}}{(k+1)!} \Gamma(1 + k\mu) \sin(\pi\mu k/2) 
                                        B\left( \frac{1}{2}, \frac{\mu k +1}{2} \right) \frac{1}{t^{\mu k}}
\label{eq:waitingTimedistrfctExp}
\end{equation}
where $B(n,m) = \Gamma(n) \Gamma(m)/\Gamma(n+m)$ is the beta function.
The expansion (\ref{eq:waitingTimedistrfctExp}) evaluated by making use of
an asymptotic expansion of the L\'{e}vy function and may be useful for numerical 
calculations of $S_T(t)$ for large values of $t$.

It is interesting to compare these results with a different assumption 
for the function $f$. Assume that it is the sum of moduli
of price returns and the waiting time that conforms to the L\'{e}vy distribution 
$f = |x_1| + \ldots + |x_p| + t$. Looking at Fig.\ref{fig:RandWalk} one could say
that it is the sum of neighboring horizontal and vertical sections $|t_n| + |x_n|$
in one knee of the random staircase
that conforms to the L\'{e}vy distribution in large time limit.
The transformation of variables (\ref{eq:mapping}) is now given by:
\begin{equation}
  \begin{array}{lll}
          x_1 & = & \theta_1 f \\
          x_2 & = & \theta_2 f \\
          \vdots & = & \vdots \\
          x_p & = & \theta_p f \\
          t   & = & f ( 1 - \theta_1 - \ldots - \theta_p ) \\ 
  \end{array}
\label{eq:trafoII}
\end{equation}
where $0 < \theta_i < 1$ and $\sum_i \theta_i \le 1$.
The Jacobian reads $(-1)^p f^p$ and the joint pdf takes the form $\phi_{\vec{X},T}(\vec{x},t) = L_{\mu,0}(f)/f^p$.
The joint pdf $\phi_{X,T}(x,t)$ which depends on the modulus of the $\vec{x}$ vector in the $L_1$ norm 
$x = \sum_{i=1}^p x_i$ reads:
\begin{eqnarray}
\phi_{X,T}(x,t) &=& \int\limits_{\sum_i x_i = x} \phi_{\vec{X},T}(\vec{x},t) d\vec{x} = 
\frac{L_{\mu,0}(x + t)}{(x + t)^p} \int\limits_{\sum_i x_i = x} d\vec{x} = 
 \nonumber \\
&=& \frac{x^{p-1} L_{\mu,0}(x + t)}{(p-1)! (x + t)^p}
\label{eq:CumulProbabI}
\end{eqnarray} 
The cumulative probability distribution of waiting times $S_T(t)$ reads:
\begin{eqnarray}
S_T(t) = \int\limits_t^\infty d\xi \int\limits_0^\infty dx \frac{x^{p-1} L_{\mu,0}(x + \xi)}{(p-1)! (x + \xi)^{p}} = 
\frac{1}{p!} \int_t^\infty L_{\mu,0}(z) \left( 1 -\frac{t}{z} \right)^p dz
\label{eq:CumulProbabII}
\end{eqnarray}
and that of price returns $S_X(x)$ takes the form:
\begin{equation}
S_X(x) = \int\limits_x^\infty d\xi \int\limits_0^\infty dt \frac{\xi^{p-1} L_{\mu,0}(\xi + t)}{(p-1)! (\xi + t)^{p}} =  \frac{1}{p!} \int_x^\infty L_{\mu,0}(z) \left( 1 -\left(\frac{x}{z}\right)^p \right) dz
\label{eq:CumulProbabIII}
\end{equation}

For both cases considered above the cumulative density of either
waiting times of price returns takes the form:
$S_W(w) = \int_w^\infty K_W(w,\xi) L_{\mu,0}(\xi) d\xi$
where $W \in \{X,T\}$ and $w \in \{x,t\}$ and the kernels $K_W(w,\xi)$
depend on the function $f(x,t)$ and on the number of stocks $p$. 
The functional form of $S_W(w)$  allows us to work out an asymptotic expansion 
of the cumulative density of waiting times
\begin{equation}
S_T(t) = \frac{1}{\pi p!} \sum_{k=1}^\infty \frac{(-1)^{k+1}}{k!} B\left(p+1, \mu k\right) \Gamma(1 + \mu k) \sin(\pi \mu k/2) \frac{1}{t^{\mu k}}
\end{equation}
and that of price returns:
\begin{equation}
S_X(x) = \frac{1}{\pi p!} \sum_{k=1}^\infty \frac{(-1)^{k+1}}{k!} \frac{p}{\mu k(\mu k+p)} \Gamma(1 + \mu k) \sin(\pi \mu k/2) \frac{1}{x^{\mu k}}
\end{equation}

The results are summarized in Table \ref{tab:pdfs}.
\begin{table}
\begin{tabular}{||l|l|l|l|l|l||} \hline
Function        &   Joint pdf         &  \multicolumn{2}{c|}{Kernels}            & \multicolumn{2}{c|}{Asymptotic behavior} \\
                &                     &  \multicolumn{2}{c|}{of cumulative pdfs} & \multicolumn{2}{c|}{of cumulative pdfs}  \\\cline{3-4} \cline{5-6}
                &                     &  in time       & in price returns   & in time & in price returns \\
$f=f(x,t)$      &   $\phi_{X,T}(x,t)$ &  $K_T(t,\xi)$  &    $K_X(x,\xi)$    & &          \\ \hline
$\sqrt{x^2+t^2}$& $\frac{L_{\mu,0}(f)}{f}$             & $\arccos{\frac{t}{\xi}}$ & $\arccos{\frac{x}{\xi}}$  
                & $C / t^\mu$  & $C / t^\mu$ \\
$x + t$         & $\frac{x^{p-1}L_{\mu,0}(f)}{f^p}$  & $(1 - \frac{t}{\xi})^p$  & $(1 - (\frac{x}{\xi})^p)$ 
                & $D/t^\mu$    & $E / t^\mu$ \\  \hline
\end{tabular}
\caption{Probability distribution functions in a CTRW with non-independent increments.
         The kernels $K_X(x, \xi)$ and $K_T(t, \xi)$ determine the cumulative densities in the following way:
         $S_W(w) = \int_w^\infty K_W(w,\xi) L_{\mu,0}(\xi) d\xi$  where $W \in \{X,T\}$ and $w \in \{x,t\}$.
         The constants $C$ and $K$ which define the asymptotic, large argument behavior read
         $C = 1/(2\pi\mu) \Gamma{(1 + \mu)} \sin((\pi \mu/2)) B(1/2, (\mu+1)/2)$, 
         $D = 1/(\pi p!) \Gamma{(1 + \mu)} \sin((\pi \mu/2)) B(p+1, \mu)$ and 
         $E = D p/\left( \mu (\mu  + p) B(p+1, \mu) \right)$
\label{tab:pdfs}} 
\end{table}

Notice that it is not possible to derive a small $w$ expansion
of $S_W(w)$ around $w=0$ even though a Taylor expansion of the L\'{e}vy function $L_{\mu,0}(x)$ around $x=0$
exists for certain values of the parameter $\mu$.
Indeed $S_W(w)$ cannot be expanded around $w=0$ because all its derivatives at $w=0$ do not exist.
Firstly, for the case $f = \sqrt{x^2 + t^2}$ we get from equation (\ref{eq:JointPDF}) 
\begin{equation}
d S_T(t)/d t \;=\; -\phi_T(t) \;=\; \int_0^\infty \phi_{X,T}(x,t) dx 
               = \int_0^\infty  \frac{2}{\pi} \frac{L_{\mu,0}(\sqrt{x^2 + t^2})}{\sqrt{x^2 + t^2}} dx
\label{eq:Derivative}
\end{equation}
and see that the integral on the right-hand side of (\ref{eq:Derivative}) diverges for $t=0$ since it contains
an non-integrable $1/x$ singularity at $x=0$. The same happens for the case in the second row in Table \ref{tab:pdfs}.

\subsection{The evolution equation \label{sec:masterEq}}
We now work out an evolution equation for $p(x,t) = P( X_t = x )$ 
ie the probability that the stochastic process $X_t$ which describes 
the time dependence of price returns has value $x$ at time $t$. 
We know [\citet{Montroll,Balescu}] that $\tilde{p}(k,s)$,
the Fourier transform of $p(x,t)$ with respect to $x$ 
and a Laplace transform with respect to time $t$, takes the form:
\begin{equation}
 \tilde{p}(k,s) = \frac{1 - \hat{\phi}(s)}{s} \cdot \frac{1}{1 - \tilde{\phi}(k,s)}
\label{eq:probability}
\end{equation}
where 
\begin{eqnarray}
 \tilde{\phi}(k,s) = F_x\left[L_t\left[ \phi_{X,T}(x,t) \right]\right] \nonumber \\
 \hat{\phi}(s)     = L_t\left[ \phi_T(t) \right]
\end{eqnarray}
Here the variables in parenthesis $(x,k)$ and $(t,s)$ are mutually conjugate
and $F_x$, $L_t$ are Fourier-Laplace transforms (FLTs) with respect to $x$ and $t$ respectively.

The evolution equation is worked out by taking inverse FLTs of (\ref{eq:probability}) 
 (compare with equation (1) in \citet{SabatKeat} and takes the form :
\begin{equation}
\int\limits_0^t \Phi(t - t') \frac{\partial}{\partial t'} p(x,t') dt' =
-p(x, t) + \int\limits_0^t \int_{-\infty}^\infty \lambda(x - x',t - t') p(x', t') dx' dt'
\label{eq:EvolutEq}
\end{equation}
where the kernel $\Phi(t)$ is related in Laplace space to the cumulative probability density 
$\hat{S}(s) = L_t\left[ S_T(t) \right]$ or to the probability density $\hat{\phi}(s)$ by:
\begin{equation}
 \hat{\Phi}(s) = \frac{\hat{S}(s) }{ 1 - s \hat{S}(s) } = \frac{1 - \hat{\phi}(s)}{s \hat{\phi}(s)}
\label{eq:kernel}
\end{equation}
and the kernel $\lambda(x,t)$ is related in Laplace-Fourier space to both the joint 
pdf and to the pdf of waiting times by:
\begin{equation}
\tilde{\lambda}(k, s) = \frac{\tilde{\phi}(k,s)}{\hat{\phi}(s)}
\label{eq:PositionKernel}
\end{equation}
In general the process is non-Markovian, ie the present value of $p(x,t)$
depends on the whole history of both $p$ and its time derivative 
$\frac{\partial}{\partial t'} p(x,t')$, unless the time dependence of the kernels reduces
to a Dirac delta function. 
To demonstrate the non-Markovian behavior for our theory
all we need to do is show that 
$\hat{\Phi}(s) \neq 1$ in (\ref{eq:kernel}) or equivalently that 
$\hat{\phi}(s) \neq 1/(1 + s)$. Then:

\begin{eqnarray}
\hat{\phi}(s) &=& \frac{2}{\pi} \int\limits_0^\infty \exp\{-s t\} \int\limits_t^\infty \frac{L_{\mu,0}(\xi)}{\sqrt{\xi^2 - t^2}} d\xi dt = \frac{2}{\pi} \int\limits_0^\infty L_{\mu,0}(\xi) d\xi \int\limits_0^\xi dt \frac{\exp\{-s t\}}{\sqrt{\xi^2 - t^2}} \label{eq:Cauchy01} \\
              &=& \frac{2}{\pi} \int\limits_0^\infty L_{\mu,0}(\xi)d\xi \int\limits_0^{\pi/2} \exp\{ -(s \xi) \sin{\theta} \} d\theta \label{eq:Cauchy02} \\
              &=& \frac{2}{\pi} \int\limits_0^{\pi/2} \int\limits_0^\infty L_{\mu,0}(\xi) \exp\{ -(s \xi) \sin{\theta} \} d\xi d\theta \label{eq:Cauchy1} \\
              &=& \frac{2}{\pi} \int\limits_0^{\pi/2} \exp\{ -(s \sin{\theta} )^\mu \} d\theta \label{eq:Cauchy2}\\
              &=& 1 + \frac{1}{\sqrt{\pi}} \sum_{n=1}^\infty \frac{(-1)^n}{n!} \frac{\Gamma\left((\mu n + 1)/2\right)}{\Gamma\left(1 + \mu n/2\right)} s^{\mu n} \neq \frac{1}{1 + s} 
\end{eqnarray}
The transition from (\ref{eq:Cauchy01}) to (\ref{eq:Cauchy02}) follows from a substitution $t = \xi \sin{\theta}$
and equality of terms (\ref{eq:Cauchy1}) and (\ref{eq:Cauchy2}) follows from the integral representation of the L\'{e}vy function
and from applying the Cauchy integral theorem.

Note that the representation of the joint pdf in Laplace-Fourier space $\tilde{\phi}(k, s)$ can also 
be calculated explicitly:
\begin{eqnarray}
\tilde{\phi}(k, s) &=& \int\limits_{-\infty}^\infty \int\limits_0^\infty 
                       \frac{2}{\pi} \exp\{ i k x - s t \} \frac{L_{\mu,0}(\sqrt{x^2 + t^2})}{\sqrt{x^2 + t^2}} dt dx \label{eq:JointPDF1} \\
                   &=& \frac{2}{\pi} \int_{0}^{\pi/2}
                                     \int_{0}^{\infty} L_{\mu,0}(r) \exp\left[ -\{s \cos{\phi} - i k \sin{\phi}\} r \right] dr
                                     d \phi   \label{eq:JointPDF2} \\
                   &=& \frac{2}{\pi} \int_0^{\pi/2} \exp\{ - (s \cos{\phi} - i k \sin{\phi})^\mu \} d\phi \label{eq:JointPDF2a} \\
                   &=& \frac{2}{\pi} \int_0^{\pi/2} \exp\{ - (s \sin{\phi} - i k \cos{\phi})^\mu \} d\phi
                   \label{eq:JointPDF3}
\end{eqnarray}
Equality (\ref{eq:JointPDF1}) = (\ref{eq:JointPDF2}) follows from going to polar coordinates
$\left( x, t\right) \longrightarrow \left( r\sin{\phi}, r\cos{\phi} \right)$
and equality  (\ref{eq:JointPDF2}) = (\ref{eq:JointPDF2a}) follows from the same transformation as that leading
from (\ref{eq:Cauchy1}) to (\ref{eq:Cauchy2}).
Finally the last equality (\ref{eq:JointPDF2a}) = (\ref{eq:JointPDF3}) follows from substitution of variables
$\phi \longleftarrow \pi/2 - \phi$. 
Equations (\ref{eq:PositionKernel}), (\ref{eq:Cauchy2}) and (\ref{eq:JointPDF3}) 
allow us to draw the conclusion that the kernel $\tilde{\lambda}(k, s)$ 
in Laplace-Fourier space does not factorize and therefore the evolution equation (\ref{eq:EvolutEq})
includes a term on the right-hand-side (the double integral over $x'$ and $t'$) that is non-local both in time and in space. 
This turns out to be different from the evolution equation in \citet{SabatKeat} where the respective term 
was non-local in space only.

Having obtained results (\ref{eq:Cauchy2}) and (\ref{eq:JointPDF3}) we can also look at the form of the evolution equation
in the limit of large price returns $x$ and large waiting times $t$.
It only amounts to finding the form of the joint pdf $\tilde{\phi}(k, s)$ and the waiting time pdf $\hat{\phi}(s)$
for small values of $k$ and $s$ and inserting them into the evolution equation (\ref{eq:probability}) 
in Laplace-Fourier space.
Expanding the exponential in (\ref{eq:Cauchy2}) in a power series and integrating term by term  we get:
\begin{equation}
\hat{\phi}(s) = 1 - \alpha_1 s^\mu + O(s^{2 \mu}) \quad\mbox{where}\quad 
\alpha_1 = B\left( \frac{\mu}{2}, \frac{1}{2} \right) \frac{1}{\pi}
\label{eq:pdfSmallExp}
\end{equation}
and from (\ref{eq:JointPDF3}) we obtain:
\begin{eqnarray}
\lefteqn{\tilde{\phi}(k, s) \simeq } \nonumber \\
\lefteqn{1 - \frac{2}{\pi} \int\limits_0^{\pi/2} \left( s \sin{\phi} - i k \cos{\phi} \right)^\mu d \phi \simeq  
         1 - k^\mu \left( \beta_1 + \beta_2 \frac{s}{k} + \beta_3 \frac{s^2}{k^2} \right)} \label{eq:jointpdfSmallExp}
\end{eqnarray}
where 
\begin{eqnarray}
\beta_1 &=& B\left( \frac{\mu + 1}{2}, \frac{1}{2} \right) \frac{e^{-i \pi/2 \mu}}{\pi} \nonumber \\
\beta_2 &=& \mu B\left( \frac{\mu}{2}, 1 \right) \frac{i e^{-i \pi/2 \mu}}{\pi} \nonumber \\
\beta_3 &=& \mu (\mu-1)/2 B\left( \frac{\mu-1}{2}, \frac{3}{2} \right) \frac{- e^{-i \pi/2 \mu}}{\pi}
\label{eq:JointPDFExp}
\end{eqnarray}
Inserting (\ref{eq:pdfSmallExp}) and (\ref{eq:jointpdfSmallExp}) into the evolution equation (\ref{eq:probability})
yields an equation
\begin{equation}
k^\mu \left( \beta_1 + \beta_2 \frac{s}{k} + \beta_3 \frac{s^2}{k^2} \right) \tilde{p}(k, s) = \alpha_1 s^{\mu - 1}
\end{equation} 
that can be reformulated in terms of fractional derivatives \citet{SamkKilb} in the $x,t$ space, 
ie after performing an inverse Laplace-Fourier transform. 
Here we merely note that the equation is different than that obtained in \citet{ScalGor}
which is not a surprise since we dropped the assumption of independence of $x$'s and $t$'s.

\section{Empirical analysis}
We now analyze two sets of stock market data, fit the theory developed
in section \ref{sec:theory} to the data and finally discuss the goodness of the fit.

The first data set comprises prices for 10 stocks(5 railways and 5 banks)
from the Irish stock exchange during the period 1850-54. 
Transactions during this period were carried on a daily basis so all waiting times are a day or longer.
Over the $5$ year period we have $1751$ data points.

The second set contains $2220695$ exchange rate quotes of US\$ to the Japanese Yen taken between 1989 to 1998 with waiting 
times ranging from a minute to some hours. The stock prices fluctuate with a much higher frequency than in the former case and
the statistical error is much smaller due to a much larger size of the data sample. This allows a more reliable fit.

\subsection{Cumulative distributions of price returns and waiting times\label{sec:CumulDistr}}
The objective is to fit theoretical cumulative density functions
$S_T(t)$ (\ref{eq:waitingTimedistrfct},\ref{eq:CumulProbabII}) and 
$S_X(x)$ (\ref{eq:CumulProbabIII})
to market data 
\begin{eqnarray}
N_T(t) &:=& \mbox{number of data points with waiting times not smaller than $t$} \nonumber \\
N_X(x) &:=& \mbox{number of data points with price returns not smaller than $x$} \nonumber
\end{eqnarray}
We consider both assumptions from Table
\ref{tab:pdfs} regarding the way returns and waiting times are correlated,
assumptions which result in two different formulas for $S_T(t)$.
However it turns out that there is no great difference between them.
Our task consists in choosing three parameters $v$, $w$ and $\mu$ in such a way that
the chi-square test statistics $\chi^2_T$ and $\chi^2_X$ are as small as possible.
\begin{eqnarray}
\chi^2_T := \sum_{t_i} \frac{( N_T(t_i)/N_T(0) - S_T(v t_i) )^2}{S_T(v t_i)} \nonumber \\
\chi^2_X := \sum_{x_i} \frac{( N_X(x_i)/N_X(0) - S_T(w x_i) )^2}{S_T(w x_i)}
\end{eqnarray}
where the sums run over the measured waiting times $t_i$ and price returns $x_i$ for $i=1,\ldots,n$.
For our data sets we chose $n=200$.
Parameter values can be estimated by fitting a power law $S_T(t) = (a_1 t)^{b_1}$ 
for $t \in (t_1, t_2)$ and $S_X(x) = (a_2 x)^{b_2}$  for $x \in (x_1, x_2)$ to the asymptotic behavior
of distribution functions. The ranges $t_1 < t < t_2$ and $x_1 < x < x_2$ have to be chosen with caution.
The variables $t$ and $x$ have to be large enough for the power law to be valid 
but cannot be too large because the finite size of the data file introduces a significant error 
to $N_T(t)$ and $N_X(x)$.

For the high frequency US\$/Yen data we have fitted following power laws to the 
distribution functions:
\begin{eqnarray}
S_T(t) &\sim& (a_1 t)^{b_1} = (0.813 t)^{-1.8662} \qquad \mbox{for $50 < t[min] < 300$} \\
S_X(x) &\sim& (a_2 x)^{b_2} = (0.451 x)^{-1.8662} \qquad \mbox{for $50 < 50000 x < 300$}  
\label{eq:USYenFits}
\end{eqnarray}
Similarly the asymptotic fits for the low frequency old Irish data read:
\begin{eqnarray}
S_T(t) &\sim& (a_1 t)^{b_1} = (1.099 t)^{-0.42} \qquad \mbox{for $30 < t[days] < 160$} \\
S_X(x) &\sim& (a_2 x)^{b_2} = (0.669 x)^{-0.42} \qquad \mbox{for $30 < 1000 x < 160$}  
\label{eq:OldIrishFits}
\end{eqnarray}
with the relative error of the coefficients not larger than a couple of percent in both cases.
It has to be emphasized that the fit to $S_X(x)$ is biased in the sense
that the power law exponent has to have the same value as in $S_T(t)$ for the theory to be valid.
Indeed, the  parameters to be fitted are expressed by the 
measured parameters $a_1,b_1$ of the asymptotic behavior in the following way 
(compare to formulas in Table \ref{tab:pdfs}):
\begin{equation}
v  = a_1 \cdot C^{1/\mu}   \qquad w  = a_2 \cdot C^{1/\mu} \qquad \mbox{and} \qquad \mu =  b_1 = b_2 
\label{eq:FitParametersI}
\end{equation}
where
$C  = 1/(2 \pi \mu) \Gamma(1 + \mu) \sin(\pi \mu/2) B\left( 1/2, (1+\mu)/2 \right)$\\
for the case $f(x, t) = \sqrt{x^2 + t^2}$ and
\begin{equation}
v  = a_1 \cdot D^{1/\mu}   \qquad w  = a_2 \cdot E^{1/\mu} \qquad \mbox{and} \qquad \mu =  b_1 = b_2 
\label{eq:FitParametersII}
\end{equation}
where
$D = 2/\pi \Gamma(1 + \mu) \sin(\pi \mu/2) B\left(\mu, P+1\right)$ and \\
$E = 2/\pi \Gamma(1 + \mu) \sin(\pi \mu/2) P/(\mu (\mu + P))$\\
for the case $f(x, t) = x + t$. \\
Recall that $B$ is the beta function defined in terms of the gamma function as 
$B(n,m) = \Gamma(n) \Gamma(m)/\Gamma(n + m)$.

\begin{figure}[!h]
\centerline{\psfig{figure=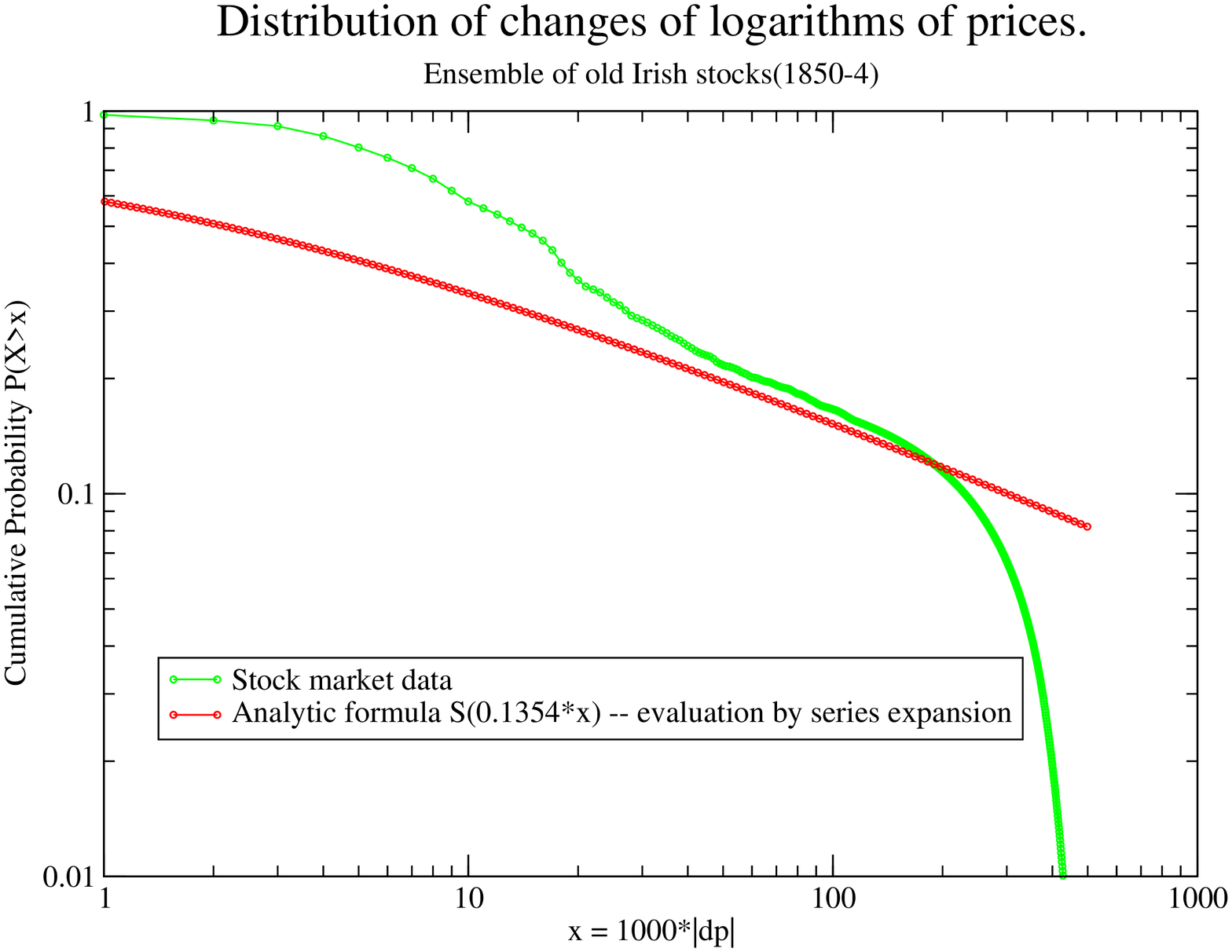,width=0.75\textwidth,angle=-90}}
\caption{The distribution of changes of logarithms of old Irish stocks
         and the theoretical curve $S_X^1(x)$.
         The test statistic equals $\chi_X^2=4.666$ for $200$ degrees of freedom.
         A large deviation at small $x$ is due to the difficulty of computing $S_X^1(x)$
         for small $x$.
         \label{fig:PriceRetI}} 
\centerline{\psfig{figure=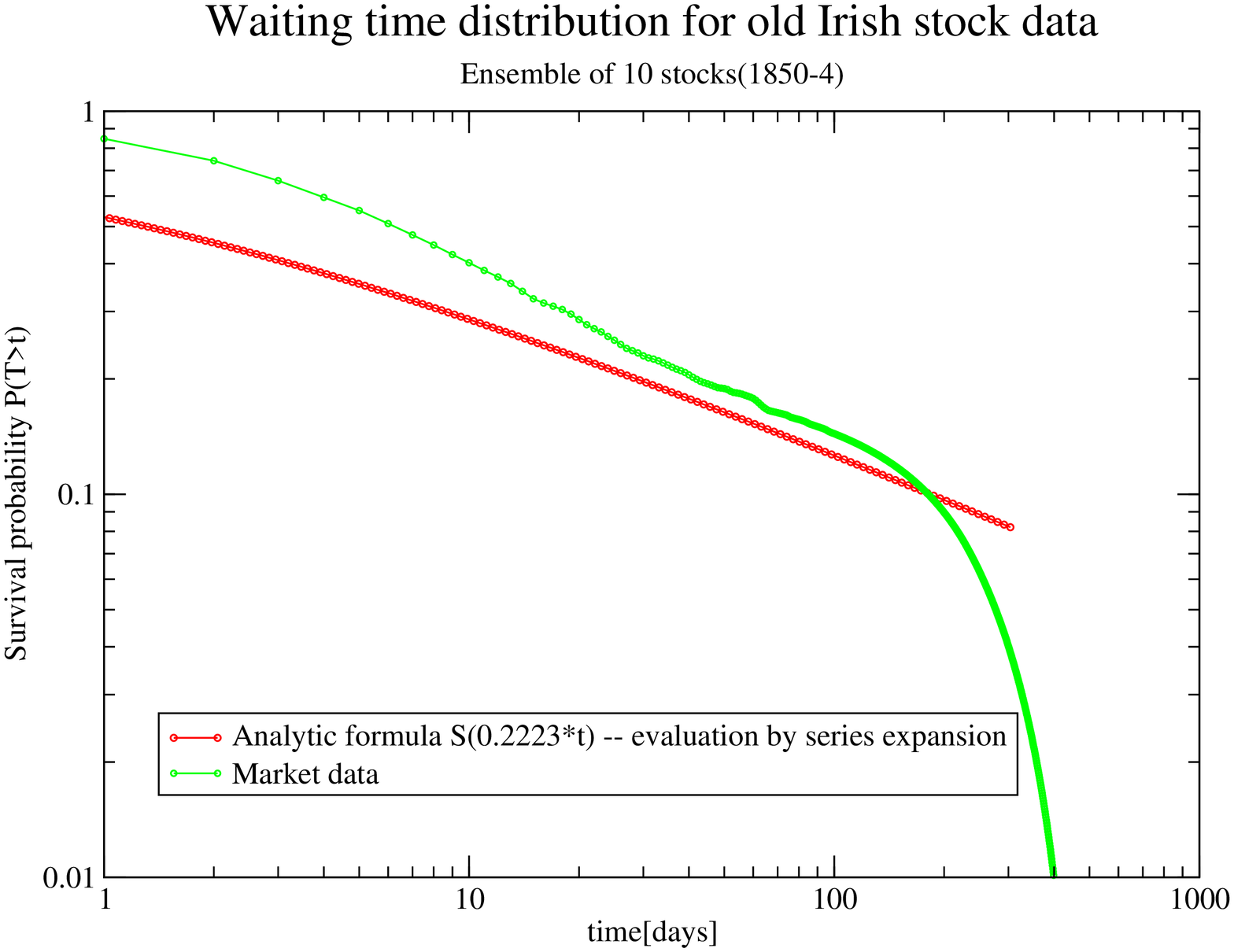,width=0.75\textwidth,angle=-90}}
\caption{Like in Fig. \ref{fig:PriceRetI} but for waiting times. 
         The test statistic equals $\chi_T^2=1.7595$ for $200$ degrees of freedom.\label{fig:WaitTimesI}} 
\end{figure}

From formulas (\ref{eq:OldIrishFits}),(\ref{eq:FitParametersI}) and (\ref{eq:FitParametersII})  we obtain
values of parameters $\mu = 1.8662$, $v = 0.1649(0.1072)$, $w = 0.0914(0.1044)$ for the US\$/Yen data
for the cases $f=\sqrt{x^2+t^2}$ and $f=x+t$ respectively.
For the old Irish data we analyzed the first case $f=\sqrt{x^2 + t^2}$ only 
and obtained the following results:
$\mu = 0.42$, $v = 0.2223$, $w = 0.1354$. 
The results are shown in Figs. \ref{fig:PriceRet},\ref{fig:WaitTimes},\ref{fig:PriceRetI},\ref{fig:WaitTimesI}.
Firstly we notice that the assumption regarding the choice of function $f(x,t)$ does not
play a significant role. The relative difference of the cumulative probability densities does not exceed $0.4$
in the US\$/Yen case. 
The price returns distribution fit is better than that of waiting times.
The $\chi^2$ test statistics $\chi^2_X = 0.497(1.974)$ are smaller than that for waiting times 
$\chi^2_T = 3.3223(2.4694)$ for the same number of degrees of freedom $304$ in both cases.
This does not however discredit the model.
The US\$/Yen waiting times data seem not to exhibit the typical features of probability distributions
of financial data, ie exponential and power law decay for small and large waiting times respectively.  
Indeed, an inspection of Fig. \ref{fig:WaitTimes} reveals  
mysterious humps at $t\simeq 100$ and $t\simeq 350$.
This could imply that 
small waiting times are
underestimated and large waiting times are overestimated.
This could be due to a limited (two decimal digits only) accuracy of the data.
Such data truncation will indeed introduce a bias toward larger waiting times.
More accurate data is required to test this suggestion.

At this point we note in more details the difficulties 
arising by computing $S_T(t)$ from formulas (\ref{eq:waitingTimedistrfct}) and (\ref{eq:waitingTimedistrfctExp}).

Since a series expansion in $t$ around $t=0$ does not exist (as proved in section \ref{sec:theory})
we computed the cumulative density $S_T(t)$ numerically
using a Gauss-Hermite quadrature. The integrand in (\ref{eq:waitingTimedistrfct})
behaves like a Gaussian for large values of $z$ and the integration range is infinite
therefore the Gauss-Hermite 
quadrature is best suited for all integration procedures.
However, since the integrand is cut off for $z < v t$ 
we use a large number 
of abscissas $z_i$ and weights $w_i$ in the quadrature
\begin{equation}
\int_0^\infty f(z) \exp(-z^2) dz = \sum_{i=1}^M f(z_i) w_i
\label{eq:GaussHermiteQuadrature}
\end{equation}
so that the contribution from the integrand for $z > v t$ is considerable.
In other words the number $M$ of abscissas has to be large enough so that the largest of abscissas
$z_{max} = max_i \; z_i$ is 
much larger than $v t$, which, for our data, is of the order of $v * 200 =  44.5$.
Furthermore since the weights $w_i$ decrease quite strongly with increasing $x_i$'s we have to 
carry out calculations with high precision.
We chose $M = 3500$ which yields $z_{max} = 83.2$ and the corresponding weight $w_{max} \sim 7 * 10^{-3011}$
and conducted computation with a precision of $300$ digits.
This method is however only suited for calculating  $S_T(t)$ for intermediate times and 
computing the cumulative density for the largest waiting times $100 < t < 500$ is marked
with a considerable error.

The Gauss-Hermite computation works quite well in the whole range of $t$ values 
for the case of US\$/Yen exchange rates, Figs. \ref{fig:PriceRet},\ref{fig:WaitTimes}.
This is because the asymptotic expansion (\ref{eq:waitingTimedistrfctExp}), 
which was derived from an asymptotic expansion of the L\'{e}vy function converges for all $t > 0$
for $ 0 < \mu < 1$ and diverges for all $t$ if $ 1 \le \mu \le 2$ \citet{MontrBendl}.
For the old Irish stock data where $\mu =0.42 < 1$ 
the asymptotic expansion (\ref{eq:waitingTimedistrfctExp}) diverges.
It is clearly difficult to work out a numerical procedure for computing
$S_T(t)$ in the whole range of $\mu$ parameter $0 < \mu < 2$.

\subsection{Goodness of the fit\label{sec:GoodFit}}
We know that the variables $z_T$ and $z_X$ conform to a $\chi^2_{n-1}$ distribution with $n-1$ degrees of freedom.
Since for large $n$ the variable $\chi^2_n$ can be approximated by
the normal distribution $Normal(n,2n)$ with mean $n$ and variance $2n$,
we see that 
for $n = 200$ the computed values 
$z_T = 2.70527$ 
$z_X = 0.910386$ 
are much smaller than the $2.5\%$-value of the distribution which is $200 + 1.96*\sqrt{400} = 239.2$,
which yields a satisfactory conformance to the theory.

\begin{figure}[!h]
\centerline{\psfig{figure=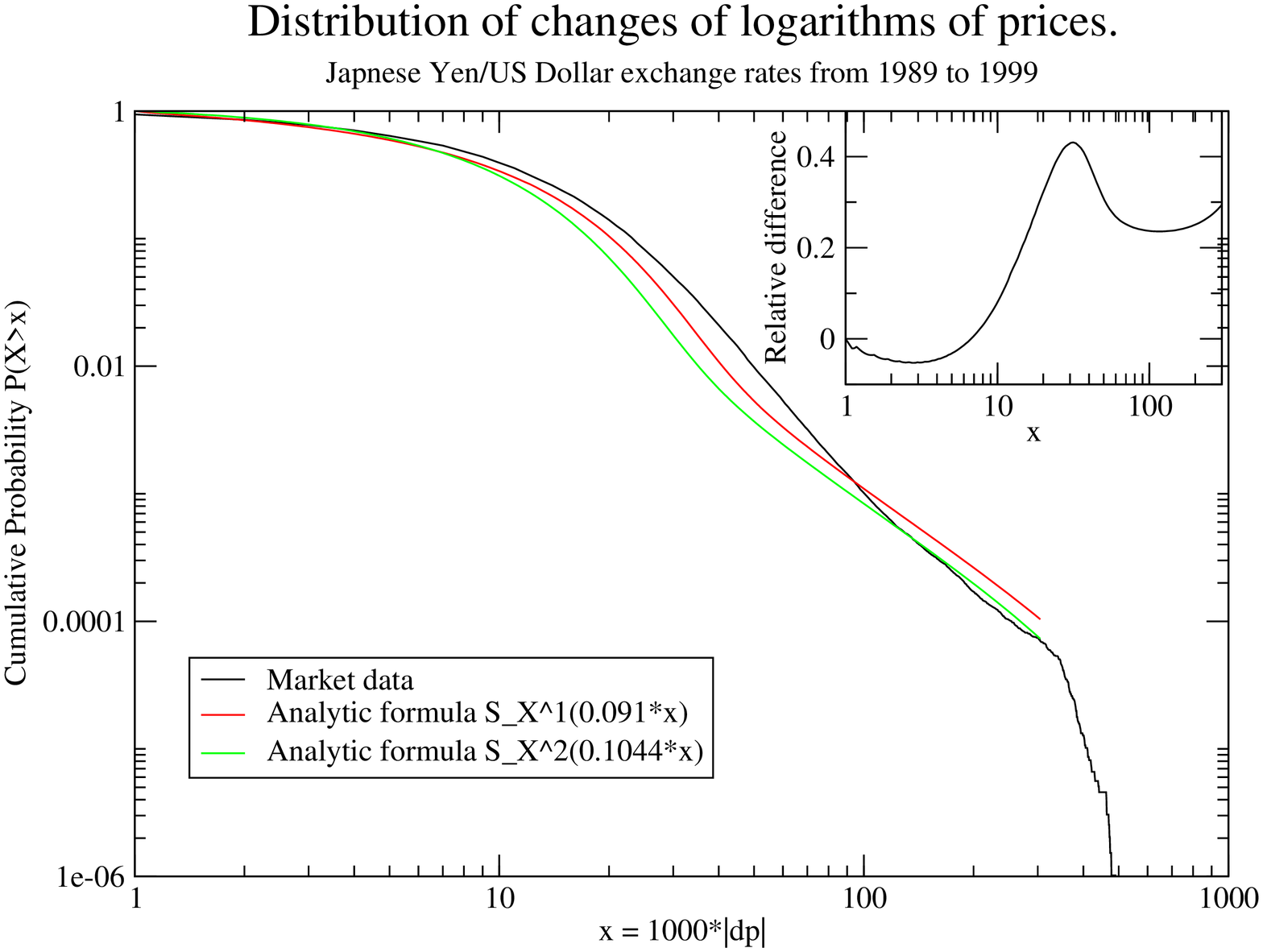,width=0.75\textwidth,angle=-90}}
\caption{The distribution of changes of logarithms of US\$/Japanese Yen exchange rates
         and the fitted theoretical formulas $S_X^1(x)$ and $S_X^2(x)$ corresponding 
         to the case in first and second row in Table \ref{tab:pdfs} respectively.
         The test statistic equals $\chi_X^2=0.497(1.974)$ for $304$ degrees of freedom.
         The inset shows the relative difference between $S_X^1$ and $S_X^2$.\label{fig:PriceRet}} 
\centerline{\psfig{figure=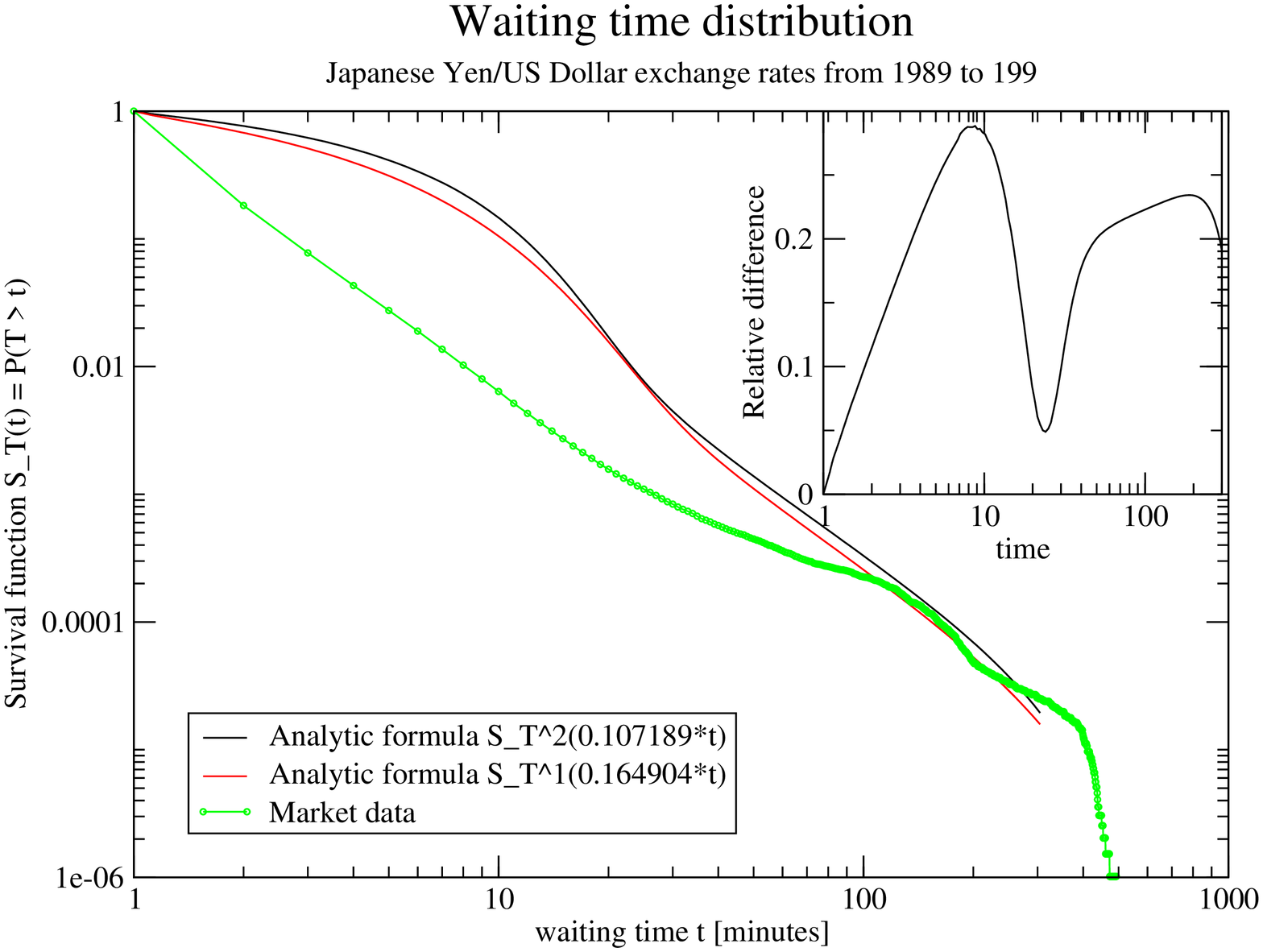,width=0.75\textwidth,angle=-90}}
\caption{Like in Fig. \ref{fig:PriceRet} but for waiting times. 
         The test statistic equals $\chi_T^2= 3.3223(2.4694)$ for $304$ degrees of freedom.\label{fig:WaitTimes}} 
\end{figure}

Here a comment has to be made on the waiting time data (Fig. \ref{fig:WaitTimes}).
Unlike the cumulative pdfs from previous papers, eg. the  
New York Stock Exchange prices investigated by \citet{RaberScal},
that are well fitted by an exponential for small values of $t$ and by a power law for large $t$ values  
the cumulative pdfs for the US\$/Yen data 
seem to exhibit a power law behavior in the whole range of $t$ values and in addition show small 
humps or shoulders for $t>100$. In this respect it is not a surprise that
the function 
(\ref{eq:waitingTimedistrfct}) does not fit the data very well and however neither did the approach 
developed in  \citet{SabatKeat}.
At least using our approach the price returns cumulative pdfs are quite well fitted 
with the same parameter $\mu$ as that used in the waiting time data.

\subsection{Correlation between price returns and waiting times \label{sec:Correl}}
We now analyze correlations between different functions
of waiting times and price returns for the market data and try to justify 
the assumption made in section \ref{sec:theory} regarding the choice of function $f(\vec{x},t)$.

The main objective of this article was to show that price returns $x$ and waiting times $t$
cannot be regarded as independent variables and to find such functions of $x$ and $t$, which,
within the accuracy of data available can be regarded as independent.

Indeed a standard contingency test between $x$ and $t$ carried out in Table \ref{tab:TR} 
shows that the null hypothesis, 
of the variables being independent can be rejected at a $1\%$ confidence level.

This begs a question whether there is a mapping of $x$ and $t$ onto new variables
that can be considered as independent. 
In section \ref{sec:theory} we considered the mapping (\ref{eq:mapping}) and we assumed
all functions $\Theta_i$ in (\ref{eq:trafo}) to be equal to one.
This is equivalent to claiming that the variables $f,\theta_1,\ldots,\theta_p$
are independent while the original $x_1,\ldots,x_p$ are not. 
Let us now check this assumption numerically making use of the market data.

Consider the case $p=1$ in (\ref{eq:mapping}), ie a case of only one stock, 
and set up a contingency table between
$r=\sqrt{(wx)^2 + (vt)^2}$ and $\phi = \arctan{(vt)/(wx)}$ in Tables \ref{tab:Rfi} and \ref{tab:Rfi1}.

The prices of stocks were given with an accuracy of two decimal digits,
which allows computation of
the error 
\begin{equation}
dx = d\log{p_{n+1}/p_n} = d p_{n+1}/p_{n+1} + d p_n/p_n \simeq 0.01/<p_n> \simeq 10^{-4}
\label{eq:LogReturnErr}
\end{equation}
From this information and from the range in which waiting times and log-price returns vary 
$1 \le t \le 2606$ and $10^{-9} \le x \le 1.9 10^{-2}$
we calculate the accuracy $r \pm dr$ and $\phi \pm d\phi$ of random variables $r=\sqrt{(wx)^2 + (vt)^2}$ 
and $\phi = \arctan{(vt)/(wx)}$, the 
mutual independence of which we are going to investigate.
Total differentials of $r$ and $\phi$ yield:
\begin{eqnarray}
 dr    = \left| \sin(\phi)     \right | v dt  +  \left| \cos(\phi)       \right| w dx \nonumber \\
 d\phi = \left| \frac{\cos(\phi)}{r} \right | v dt  +  \left| \frac{\sin(\phi)}{r} \right| w dx 
\label{eq:dRdPhi}
\end{eqnarray}
We calculate the differentials $dr$ and $d\phi$ for $r,\phi$ values from Table (\ref{tab:Rfi})
and estimate the errors $\delta r$ and $\delta \phi$ by the average value of the corresponding differential.
This yields a result $\delta r = 0.165$ and $\delta\phi = 0.002$.
If we compare this with the range of values in each box of the Table (\ref{tab:Rfi}),
we get $r_{m+1}-r_m = 0.206$ and $\phi_{n+1}-\phi_n = \pi/2 0.0008 = 0.0012$ which is on the verge of 
the accuracy of measurements.

Since the Table (\ref{tab:Rfi}) contains many zeros in the upper and lower right corners it would be reasonable to 
refine the range of variability, introducing more rows in the center of the table where most data points are gathered
and assemble a couple of top rows and a couple of bottom rows into single rows.
This would however yield a refinement much below the measurement error 
unless the number of rows and thus the number of degrees of freedom was decreased resulting in a less certain estimate.  

Indeed, a contingency test in Table (\ref{tab:Rfi1}) with the same number of rows and columns but with a refined range of the variable 
$\phi = \pi/2 (1 + 8 (n-5) 10^{-4})$ yields a much larger value of the test statistic $z = 1507.7$ 
than that from Table (\ref{tab:Rfi}) suggesting the null hypothesis should be rejected.
However, since the $\phi$ values are marked with an error $\delta\phi=0.002$, 
which is one order of magnitude larger than the increment 
$\phi_{n+1}-\phi_n = \pi/2 \cdot 8 \cdot 10^{-4}$ such a test 
is not reliable.

The accuracy of measurements is not an issue for a contingency test between the log-returns $x$ and the waiting times $t$ 
themselves. 

\section{Conclusions}
We have developed a new generalization of the continuous time random walk that recognizes
waiting times and associated jumps may not be independent. The theory has been applied to stock price data.
Two data sets were used namely old $19$th century Irish daily data 
and $20$th century US\$/Japanese Yen currency data.
Since our theory avails of analytical results it would be interesting to pursue further
investigation of correlations between waiting times $T$ and price jumps $X$.
Note that averages of the kind $E\left[ X^n T^m \right]$ are readily expressed
through derivatives of the joint pdf in Fourier-Laplace space $\partial_k^n \partial_s^m\tilde{\phi}(k, s)$.
In particular, we see from (\ref{eq:JointPDFExp}) that 
$E\left[ X^\mu \right] \simeq \beta_1$, $E\left[X^{\mu-1} T \right] \simeq \beta_2$
and $E\left[X^{\mu-2} T^2 \right] \simeq \beta_3$.
Once more accurate data was available these results can be used to validate the theory.

\section{Acknowledgments}
Przemys\mbox{\l}aw Repetowicz acknowledges support by the European Commission via a Marie-Curie
Development Host Fellowship (contract number HPMD-CT-2000-00048).

\begin{table}
{\tiny
\begin{tabular}{llllll|l} \hline
$t = n*10 $ & \multicolumn{6}{l}{$x = -0.005+ 0.0025*m$} \\ \cline{2-7}
 [min]          &    $\le$ 1     &    1--2        &    2--3        &    3--4        &    $\ge$ 4     & $\sum_x$ \\ \hline
 0 --  1        &   439(560.569) &1103725(1.10344e+06) &1102567(1.10263e+06) &   364(442.292) &    91(111.318) &         2207186\\ 
 1 --  2        &    19(  2.579) &  4918(5078.31) &  5209(5074.56) &    10(  2.035) &     2(  0.512) &           10158\\ 
 2 --  3        &     9(  0.398) &   753(784.893) &   802(784.313) &     5(  0.314) &     1(  0.079) &            1570\\ 
 3 --  4        &     7(  0.141) &   294(279.462) &   251(279.255) &     5(  0.112) &     2(  0.028) &             559\\ 
 4 --  5        &    11(  0.066) &   112(130.982) &   130(130.885) &     9(  0.052) &     0(  0.013) &             262\\ 
 5 --  6        &     4(  0.046) &    88( 91.987) &    86( 91.919) &     4(  0.036) &     2(  0.009) &             184\\ 
 6 --  7        &     5(  0.032) &    63( 62.991) &    54( 62.944) &     4(  0.025) &     0(  0.006) &             126\\ 
 7 --  8        &     0(  0.017) &    28( 33.995) &    36(  33.97) &     4(  0.013) &     0(  0.003) &              68\\ 
 8 --  9        &     3(  0.011) &    15( 21.997) &    22(  21.98) &     4(  0.008) &     0(  0.002) &              44\\ 
 9 -- 10        &     1(  0.013) &    20( 26.496) &    29( 26.476) &     2(   0.01) &     1(  0.002) &              53\\ 
 $\ge$ 10       &    66(  0.123) &   181(242.467) &   191(242.288) &    34(  0.097) &    13(  0.024) &             485\\ \hline
 $\sum_t$       &            564 &        1110197 &        1109377 &            445 &            112 & 2220695
\end{tabular}
\caption{Contingency table between $x= \log(p_{n+1}/p_n)$ and $t$ for the US\$/Yen data.
          The numbers in brackets are the expected frequencies under the assumption that $x$ and $t$ are independent variables. 
          The test statistics $z = \sum_{i} (O_i - E_i)^2/E_i  \simeq \chi^2_{40} = 65619.1$
          is clearly above the $2.5\%$--value of the distribution which is 
          $\chi^2_{40,2.5\%} = 40 + 1.906*\sqrt{80} = 57.0478$ \label{tab:TR}}
\begin{tabular}{llllll|l} \hline
$t = n*10 $ & \multicolumn{6}{l}{$r = -0.125 + 0.0625*m$} \\ \cline{2-7}
 [days]         &    $\le$ 1        &    1--2        &    2--3        &    3--4        &    $\ge$ 4  & $\sum_x$ \\ \hline
 0 --  1        &     5( 17.084) &   469(482.644) &   738(697.627) &     6( 14.949) &     0(  5.694) &            1218\\ 
 1 --  2        &     2(  3.394) &   105( 95.894) &   130(138.609) &     4(   2.97) &     1(  1.131) &             242\\ 
 2 --  3        &     5(  1.683) &    49( 47.551) &    58( 68.731) &     6(  1.472) &     2(  0.561) &             120\\ 
 3 --  4        &     3(  0.673) &    17(  19.02) &    23( 27.492) &     4(  0.589) &     1(  0.224) &              48\\ 
 4 --  5        &     1(  0.378) &    12( 10.699) &    13( 15.464) &     1(  0.331) &     0(  0.126) &              27\\ 
 5 --  6        &     0(  0.182) &     8(  5.151) &     4(  7.445) &     0(  0.159) &     1(   0.06) &              13\\ 
 6 --  7        &     2(  0.154) &     5(  4.358) &     3(    6.3) &     0(  0.135) &     1(  0.051) &              11\\ 
 7 --  8        &     1(  0.098) &     2(  2.773) &     4(  4.009) &     0(  0.085) &     0(  0.032) &               7\\ 
 8 --  9        &     1(  0.098) &     4(  2.773) &     2(  4.009) &     0(  0.085) &     0(  0.032) &               7\\ 
 9 -- 10        &     0(  0.056) &     2(  1.585) &     1(  2.291) &     0(  0.049) &     1(  0.018) &               4\\ 
 $\ge$ 10       &     4(  0.196) &     5(  5.547) &     4(  8.018) &     0(  0.171) &     1(  0.065) &              14\\ \hline
 $\sum_t$       &             24 &            678 &            980 &             21 &              8 &   1711
\end{tabular}
\caption{The same as in Table \ref{tab:TR} for the old Irish data.
          The test statistics $z = 304.798$
          is clearly above the $2.5\%$--value of the distribution which is 
          $\chi^2_{40,2.5\%} = 40 + 1.906*\sqrt{80} = 57.05$ \label{tab:TR_Irish}}
}
\end{table}

\begin{table}
{\tiny
\begin{tabular}{llllll|l} \hline
$\phi =$ & \multicolumn{6}{l}{$r = 0.206*m$} \\ \cline{2-7}
 $\frac{\pi}{2}(1 + 8 (n-5) 10^{-4})$              &    $\le$ 1        &    1--2        &    2--3        &    3--4        &    $\ge$ 4 & $\sum_{r}$\\ \hline
 0 --  1        &     6(  4.937) &     0(  0.612) &     0(  0.201) &     0(   0.14) &     0(  0.108) &                6\\ 
 1 --  2        &    17( 13.988) &     0(  1.735) &     0(  0.569) &     0(  0.397) &     0(  0.308) &               17\\ 
 2 --  3        &    33( 27.154) &     0(  3.369) &     0(  1.105) &     0(   0.77) &     0(  0.599) &               33\\ 
 3 --  4        &   337(307.752) &    29( 38.188) &     7(  12.53) &     0(  8.736) &     1(  6.792) &              374\\ 
 4 --  5        &910745( 912974) &114348( 113289) & 37462(37172.2) & 26331(25917.3) & 20618(20151.2) &          1109504\\ 
 5 --  6        &915832( 913687) &112353( 113378) & 36929(37201.3) & 25542(25937.5) & 19714(20166.9) &          1110370\\ 
 6 --  7        &   311(274.837) &    19( 34.104) &     3(  11.19) &     1(  7.802) &     0(  6.066) &              334\\ 
 7 --  8        &    37( 32.091) &     2(  3.982) &     0(  1.306) &     0(  0.911) &     0(  0.708) &               39\\ 
 8 --  9        &    16( 13.165) &     0(  1.633) &     0(  0.536) &     0(  0.373) &     0(   0.29) &               16\\ 
 9 -- 10        &     0(      0) &     0(      0) &     0(      0) &     0(      0) &     0(      0) &                0\\ 
 $\ge$ 10       &     2(  1.645) &     0(  0.204) &     0(  0.067) &     0(  0.046) &     0(  0.036) &                2\\ \hline
$\sum_{t}$      &        1827336 &         226751 &          74401 &          51874 &          40333 & 2220695
\end{tabular}
\caption{Contingency table between $r = \sqrt{(wx)^2 + (vt)^2}$ and $\phi = \arctan{(vt)/(wx)}$ for the US \$/Yen data.
         The numbers in brackets are the expected frequencies under the assumption that $r$ and $\phi$ are independent variables. 
         The null hypothesis cannot stand but the test statistics 
         $z = \sum_{i} (O_i - E_i)^2/E_i  \simeq \chi^2_{40} = 138.646 > 40 + 1.906*\sqrt{80} = 57.048$
         has a much smaller value than in Table 1
         \label{tab:Rfi}}
\begin{tabular}{llllll|l} \hline
$\phi = \frac{\pi}{2}(1 + (n-5) 4 10^{-4})$ & \multicolumn{6}{l}{$r = 0.206*m$} \\ \cline{2-7}
                &    $\le$ 1     &    1--2        &    2--3        &    3--4        &    $\ge$ 4     & $\sum_r$ \\ \hline
 0 --  1        &    56(  46.08) &     0(  5.718) &     0(  1.876) &     0(  1.308) &     0(  1.017) &              56\\ 
 1 --  2        &    68( 66.652) &    13(   8.27) &     0(  2.713) &     0(  1.892) &     0(  1.471) &              81\\ 
 2 --  3        &   269(241.099) &    16( 29.917) &     7(  9.816) &     0(  6.844) &     1(  5.321) &             293\\ 
 3 --  4        &  4071(3419.83) &    71(424.361) &     6( 139.24) &     5( 97.081) &     3( 75.482) &            4156\\ 
 4 --  5        &906674( 909554) &114277( 112865) & 37456(  37033) & 26326(25820.2) & 20615(20075.7) &         1105348\\ 
 5 --  6        &911819( 910295) &112267( 112957) & 36920(37063.2) & 25534(25841.3) & 19709(20092.1) &         1106249\\ 
 6 --  7        &  4013(3391.03) &    86(420.787) &     9(138.067) &     8( 96.263) &     5( 74.846) &            4121\\ 
 7 --  8        &   238(207.362) &    10( 25.731) &     3(  8.442) &     1(  5.886) &     0(  4.576) &             252\\ 
 8 --  9        &    73( 67.475) &     9(  8.372) &     0(  2.747) &     0(  1.915) &     0(  1.489) &              82\\ 
 9 -- 10        &    28(  23.04) &     0(  2.859) &     0(  0.938) &     0(  0.654) &     0(  0.508) &              28\\ 
 $\ge$ 10       &    27( 23.863) &     2(  2.961) &     0(  0.971) &     0(  0.677) &     0(  0.526) &              29\\ \hline
 $\sum_t$       &        1827336 &         226751 &          74401 &          51874 &          40333 & 2220695
\end{tabular}
\caption{The same as in Table \ref{tab:Rfi} only for a refined range in $\phi$.
         Now the test statistic is larger than before $z = 1507.75$ but the amount by which $\phi$ increases
         $d\phi \simeq 4*10^{-4}$ is on the verge of the measurement accuracy. \label{tab:Rfi1}}
}
\end{table}

%





\begin{thebibliography}{99}


\bibitem[Scalas(2000)]{ScalGor} 
  E.Scalas et al., Fractional calculus and continuous-time finance, Physica A 284 (2000) 376
\bibitem[Mainardi(2000)]{MainRaber}
  F.Mainardi et al.,  Fractional calculus and continuous-time finance II: the waiting time distribution, Physica A 287 (2000) 468--481
\bibitem[Sabatelli(2002)]{SabatKeat} 
 L.Sabatelli et al., Waiting time distributions in financial markets, Eur. Phys. J. B {\bf 27} (2002) 273--275

\bibitem[Raberto(2002)]{RaberScal} M.Raberto et al., Waiting times and returns in high-frequency financial data: an empirical study, Physica A 314 (2002) 749--755

\bibitem[Bachelier(1900)]{Bach} Bachelier L., Theory of Speculation, Ann. Sci. Ecole Norm. Sup. {\bf 3}, 21 (1900);
reprint from P.H. Cootner (editor), The random character of stock prices, second edition (MIT Press Cambridge, 1969)

\bibitem[Montroll(1965)]{Montr}E. Montroll and G. H. Weiss, Random Walks on Lattices. II, J. Math. Phys. {\bf 6} No. 2 (1965) 167--181)

\bibitem[Black et al.(1973)]{Black}Black F. and Scholes M., The Pricing of Options and Corporate Liabilities, Journal of Political Economy {\bf 81}, (1973) 637--659

\bibitem[Lux(1996)]{Lux}Lux T., The stable Paretian hypothesis and the frequency of large returns: an examination of major German stocks, Applied Financial Economics {\bf 6}, (1996) 463--475

\bibitem[Gopikrishnan(1999)]{GopiPle}Gopikrishnan P. et al., Scaling of distribution of fluctuations of financial market indices, Phys. Rev. E {\bf 60}, (1999) 5305--5316

\bibitem[Plerou(1999)]{PleGopi}Plerou V. et al., Scaling of distribution of price fluctuations of individual companies, Phys. Rev. E {\bf 60}, (1999) 6519--6529

\bibitem[Mandelbrot(1963)]{Mandel}Mandelbrot B., The variations of certain speculative prices, J. of Business {\bf 36} (1963) 392--417 

\bibitem[Fama(1970)]{Fama}Fama E.F., Efficient Capital Markets: A Review of Theory and Empirical Work, J. of Finance {\bf 25} (1970) 383--417 

\bibitem[Bouchaud(1990)]{Bouch}Bouchaud J.P. et al., Anomalous diffusion in disordered media, Physics Reports {\bf 195} (1990) 127--293 

\bibitem[Gopikrishnan(1998)]{GopiMeyer}Gopikrishnan P. et al., Inverse cubic law for the distribution of stock price variations, Eur. Phys. J. B {\bf 3} (1998) 139--140 

\bibitem[Montroll(1965)]{Montroll}Montroll E.W. et al., Random Walks on Lattices II, J. Math. Phys. {\bf 6} (1965) 167--181 

\bibitem[Balescu(1997)]{Balescu}Balescu R., Continuous Time Random Walks, in: {\it Statistical Dynamics Matter out of Equilibrium} Imperial College Press, World Scientific, London 1997 

\bibitem[Montroll(1984)]{MontrBendl}Montroll E.W. et al., On L\'{e}vy (or Stable) Distributions and Williams-Watts Model of Dielectric Relaxation, J. of Stat. Phys. {\bf 34} Nos. 1/2 (1984) 129--161 

\bibitem[Levy(1937)]{Levy}L\'{e}vy P., Th\'{e}orie de l'addition des variables al\'{e}atoires (Gauthier-Villars, Paris, 1937)

\bibitem[Gnedenko(1954)]{Gnedenko}Gnedenko B.V. and Kolmogorov A.N., Limit Distributions for Sums of Independent Random Variables (Addison-Wesley, Cambridge, Massachusetts, 1954)

\bibitem[Samko(1993)]{SamkKilb} Samko S.G., Kilbas A.A., Marichev O.I.,
Fractional Integrals and Derivatives, Theory and Applications,
Gordon and Breach Science Publishers, London, 1993           

\bibitem[Arfken(1970)]{Arfken}Arfken G., Mathematical Methods for Physicists (Academic Press New York and London , 1970)
\end{thebibliography}
\end{document}